\documentclass[11pt]{article}
\usepackage{amssymb}
\usepackage{graphicx}
\usepackage{lscape}
\usepackage{theorem}

\newcommand \be  {\begin{equation}}
\newcommand \bea {\begin{eqnarray}}
\newcommand \ee  {\end{equation}}
\newcommand \eea {\end{eqnarray}}
\renewcommand \epsilon {\varepsilon}

\begin{document}

\title{Hedging Extreme Co-Movements
\footnote{We acknowledge helpful discussions and
exchanges with J.P. Laurent. This work was partially
supported by
the James S. Mc Donnell Foundation 21st century scientist award/studying
complex system.}}
\author{Y. Malevergne$^{1,2}$ and D. Sornette$^{1,3}$ \\
\\
$^1$ Laboratoire de Physique de la Mati\`ere Condens\'ee CNRS UMR 6622\\
Universit\'e de Nice-Sophia Antipolis, 06108 Nice Cedex 2, France\\
$^2$ Institut de Science Financi\`ere et d'Assurances - Universit\'e Lyon I\\
43, Bd du 11 Novembre 1918, 69622 Villeurbanne Cedex, France\\
$^3$ Institute of Geophysics and Planetary Physics
and Department of Earth and Space Science\\
University of California, Los Angeles, California 90095, USA\\
\\
email: Yannick.Malevergne@unice.fr and sornette@unice.fr\\
fax: (33) 4 92 07 67 54\\
}

\maketitle

{\bf  Based on a recent theorem due to the authors, it is shown
how the extreme tail
dependence between an asset and a factor or index or between two assets
can be easily calibrated. Portfolios constructed with stocks with minimal
tail dependence with the market exhibit a remarkable degree of decorrelation
with the market at no cost in terms of performance measured by the
Sharpe ratio.}

\vskip 1cm

Over a hundred years ago,
Vilfred Pareto
discovered a statistical relationship, now known as the 80-20 rule,
which manifests itself over and over in large systems:
``In any series of elements to be controlled, a
selected small fraction, in terms of numbers of
elements, always accounts for a large fraction in terms of effect.''
The stock market is no exception: events occurring over a very small fraction
of the total invested time may account for
most of the gains and/or losses. Diversifying away such large risks
requires novel approaches to portfolio management, which must take
into account the non-Gaussian fat tail structure of distributions
of returns and their dependence. Recent shocks and crashes have shown that
standard portfolio diversification work well in normal times but may
break down in stressful times, precisely when diversification is
the most important: as a caricature, one could say that diversification
works when one does not really need it and may fail severely when it is
most needed.

Technically, the question boils down to whether large price movements
occur mainly in an isolated manner or in a coordinated way. This question is
vital for fund managers who take advantage of the diversification to
hedge their risks. Here, we introduce a new technique to quantify
and empirically estimate
the propensity for assets to exhibit extreme co-movements, through the
use of the so-called {\it coefficient of tail
dependence}. Using a factor model framework and tools from extreme
value theory,
we provide novel analytical formulas for the coefficient of tail
dependence between
arbitrary assets, which yields an efficient non-parametric estimator.
We then construct portfolios of stocks with minimal tail dependence
with the market
represented by the Standard and Poor's 500 index and show that their superior
behavior in stressed times come together with qualities in terms
of Sharpe ratio and standard quality measures that are not inferior to
standard portfolios.

\section{Assessing large co-movements}

Standard estimators of the dependence between
assets are the correlation coefficient or the Spearman's
rank correlation for instance. However, as stressed by \cite{EMS99},
these kind of dependence measures suffer from many deficiencies.
Moreoever, their values are
mostly controlled by relatively small moves
of the asset prices around their mean. To cure
this problem, it has been proposed to use the correlation
coefficients conditioned on large movements of the assets. But
\cite{Boyer_etal} have emphasized that this approach suffers also from a severe
systematic bias leading to spurious strategies:
the conditional correlation in general evolves with
time even when the true non-conditional correlation remains constant.
In fact, \cite{MS02a} have
shown that any approach based on conditional dependence measures
implies a spurious change of the intrinsic value of the dependence, measured
for instance by copulas. Recall that the copula of several random variables is
the (unique) function which completely embodies the dependence between these
variables, irrespective of their marginal behavior (see \cite{Nelsen}
for a mathematical description of the notion of copula).

In view of these limitations of the
standard statistical tools, it is natural to turn to
extreme value theory. In the univariate case,
extreme value theory is very
useful and provides many tools for investigating the extreme tails of
distributions of assets returns. These new developments rest on
the existence of a few fundamental results on extremes, such as the
Gnedenko-Pickands-Balkema-de Haan theorem which gives a general
expression for the distribution of exceedance over a large
threshold. In this framework,
the study of large and extreme co-movements
requires the multivariate extreme values theory, which unfortunately
does not provide strong results. Indeed, in constrast with the
univariate case, the class of limiting extreme-value
distributions is too broad and cannot be used to
constrain accurately the distribution of large co-movements.

In the spirit of the mean-variance portfolio or of utility theory
which establish an investment decision on a unique risk measure, we use the
{\it coefficient of tail dependence}, which, to our knowledge,
was first introduced in the financial context by \cite{EMS01}.
The coefficient of tail dependence between assets $X_i$ and $X_j$
is a very natural and easy to
understand measure of extreme co-movements. It is defined as
the probability that the
asset $X_i$ incurs a large loss (or gain) assuming that the asset
$X_j$ also undergoes a large loss (or gain) at the same probability
level, in the limit where this probability level explores the extreme tails
of the distribution of returns of the two assets.
Mathematically speaking, the coefficient of {\it lower} tail
dependence between the two assets $X_i$ and $X_j$, denoted by
$\lambda_{ij}^-$ is defined by
\be
\label{eq:l-}
\lambda_{ij}^- = \lim_{u \rightarrow 0} \Pr \{X_i < {F_i}^{-1}(u)~|~ X_j
< {F_j}^{-1}(u)
\},
\ee
where ${F_i}^{-1}(u)$ and ${F_j}^{-1}(u)$ represent the quantiles of assets
$X_i$ and $X_j$ at the level $u$. Similarly, the coefficient of {\it upper}
tail dependence is
\be
\label{eq:l+}
\lambda_{ij}^+ = \lim_{u \rightarrow 1} \Pr \{X_i > {F_i}^{-1}(u)~|~ X_j
>  {F_j}^{-1}(u)
\}.
\ee
$\lambda_{ij}^-$ (respectively $\lambda_{ij}^+$) is of concern to investors
with long (respectively short) positions. We refer to
\cite{CHT99} and references therein for a survey of the
properties of the coefficient of tail dependence. Let us stress
that the use of quantiles in the definition of $\lambda_{ij}^-$
and $\lambda_{ij}^+$ makes them independent of the
marginal distribution of the asset returns: as a consequence, the
tail dependence parameters are intrinsic dependence measures.
The obvious gain is an ``orthogonal'' decomposition
of the risks into (1) individual risks carried by
the marginal distribution of each asset and (2) their collective risk
described by their dependence structure or copula.

Being a probability, the coefficient of tail dependence varies
between $0$ and $1$. A large value of $\lambda_{ij}^-$ means that
large losses occur almost surely together. Then, large risks
can not be diversified away and the assets crash together.
This investor and portfolio manager nightmare is further amplified
in real life situations by the limited liquidity of markets.
When $\lambda_{ij}^-$ vanishes, these
assets are said to be asymptotically independent, but this term
hides the subtlety that the assets can still present a non-zero
dependence in their tails. For
instance, two normally distributed assets can be shown
to have a vanishing coefficient
of tail dependence. Nevertheless, unless their correlation coefficient
is identically zero,
these assets are never independent. Thus, asymptotic independence must
be understood as the weakest dependence which can be quantified by the
coefficient of tail dependence (for other details, the reader is
refered to \cite{LT98}).

For practical implementations, a direct application of the
definitions (\ref{eq:l-}) and (\ref{eq:l+}) fails to provide
reasonable estimations due to the double curse of dimensionality and
undersampling of extreme values,
so that a fully non-parametric approach is not
reliable. It turns out to be possible to circumvent this
fundamental difficulty by considering the general class of
factor models, which are among
the most widespread and versatile models in
finance. They come in two classes: multiplicative
and additive factor models respectively. The multiplicative
factor models are generally
used to model asset fluctuations due to an underlying
stochastic volatility (see for instance \cite{HW} and \cite{T94} for a
survey of the properties of these models). The additive
factor models are made to relate asset
fluctuations to market fluctuations, as in the Capital
Asset Princing Theory (CAPM) and its generalizations (see \cite{S64,R73} for
instance), or to any set of common factors as in \cite{R76}'s
Arbitrage Pricing Theory. The coefficient of tail dependence is known
in close form for both classes of factor models, which allows, as
we shall see, for an efficient empirical estimation.

\section{Tail dependence generated by factor models}

We first examine multiplicative factor models
which account for most of the stylized facts observed on financial time
series. Basically, a multivariate stochastic volatility model with a commom
stochastic volatility factor can be written as
\be
{\bf X} = \sigma \cdot {\bf Y}~,
\ee
where $\sigma$ is a positive random variable modeling the volatility,
${\bf Y}$ is a Gaussian random vector and ${\bf X}$ is the vector of
assets returns. In this framework, the multivariate distribution of
assets return ${\bf X}$ is an elliptical multivariate
distribution. For instance, if the inverse of the square of the volatility
follows the $\chi^2$-distribution with $\nu$ degrees of freedom, the
distribution of assets return will be the Student's distribution with $\nu$
degrees of freedom. When the volatility follows ARCH or GARCH
processes, then the assets returns are also elliptically distributed
with fat tailed marginal distributions. Thus, any asset $X_i$ is
asymptotically distributed according to a power law\footnote{More
precisely, the $X_i$'s follow regularly varying distributions (see
\cite{BGT97} for details on regular variations).} : $\Pr\{|X_i| > x \}
\sim x^{-\nu}$, with the same exponent $\nu$ for all assets, due to
the ellipticity of their multivariate distribution.

\cite{HL01} have shown that the necessary and sufficient condition for
any two assets $X_i$ and $X_j$ to have a non-vanishing coefficient of
tail dependence is that their distribution be regularly varying.
Denoting by $\rho_{ij}$ the correlation coefficient
between the assets $X_i$ and $X_j$ and by $\nu$ the tail index of their
distributions, they obtain:
\be
\label{eq:lambda_mul}
\lambda_{ij}^\pm= \frac{\int_{\pi/2-\arcsin \rho_{ij}}^{\pi/2}~dt ~\cos^\nu
   t}{\int_0^{\pi/2}~dt ~\cos^\nu t} = 2~ I_\frac{1+\rho_{ij}}{2} \left(
\frac{\nu+1}{2}, \frac{1}{2} \right)~,
\ee
where $I_\frac{1+\rho_{ij}}{2} (x,y)$ denotes the incomplete gamma function.
This expression holds for any regularly varying elliptical
distribution, irrespective of the exact shape of the
distribution. Only the tail index is important in the determination of
the coefficient of tail dependence because $\lambda_{ij}^\pm$ probes
the extreme
end of the tail of the distributions which all have roughtly speaking the same
behavior for regularly varying distributions.
In constrast, when the marginal distributions decay faster than any
power law, such as for the Gaussian,
exponential and gamma distributions, the
coefficient of tail dependence is zero.

Let us now turn to the second class of additive factor models,
whose introduction in finance goes back at least to
the Arbitrage Pricing Theory \cite{R76}. They are now widely used
in many branches of finance, including to model stock returns, interest
rates and credit risks. Here, we shall only consider the effect of
a single factor, which may represent the
market for instance. This factor will be denoted by $Y$ and its
distribution by $F_Y$. As previously, the vector ${\bf X}$ is the vector
of assets returns and ${\bf \epsilon}$ will denote the vector of
idiosyncratic noises assumed independent\footnote{In fact $\epsilon$
   and $Y$ can be weakly dependent. See \cite{MS02b} for details.} of
$Y$. ${\bf \beta}$ is the
vector whose components are the regression coefficients of the $X_i$
on the factor $Y$. Thus, the factor model reads:
\be
{\bf X} = {\bf \beta} \cdot Y + {\bf \epsilon}~.
\label{idio}
\ee
In constrast with multiplicative factor models, the multivariate
distribution of ${\bf X}$ cannot be
obtained in an analytical form, in the general case. Obviously,
when $Y$ and ${\bf \epsilon}$ are normally distributed, the
multivariate distribution of ${\bf X}$ is also normal
but this case is not very interesting. In a sense,
additive factor models are richer than the multiplicative ones, since
they give birth to a larger set of distributions of assets returns.

Notwithstanding these difficulties, it turns out to be possible to
obtain the coefficient of tail
dependence for any pair of assets $X_i$ and $X_j$. In a first step,
let us consider the coefficient of tail dependence
$\lambda_i^\pm$ between any asset $X_i$ and the factor $Y$ itself.
\cite{MS02b} have shown that $\lambda_i^\pm$ is also identically zero
for all rapidly
varying factors, that is, for all factors whose distribution decays
faster than any power law, such as the Gaussian, exponential
or gamma laws. When the factor $Y$ has a distribution which decays
regularly with tail
index $\nu$, we have
\be
\label{eq:lambda_add}
\lambda_i^+ = \frac{1}{\max\left\{1, \frac{l}{\beta_i}\right\}^\nu},
~~~~~~~ \mbox{\rm where}~~~~~~~ l= \lim_{u \rightarrow 1}
\frac{F_X^{-1}(u)}{F_Y^{-1}(u)}.
\ee
A similar expression holds for $\lambda_i^-$ obtained by simply
changing the limit $u \to 1$ by $u \to 0$ in the definition of $l$.
$\lambda_i^\pm$ is non-zero as long as
$l$ remains finite, that is, when the tail
of the distribution of the factor is not thinner than the tail of the
idiosyncratic noise $\epsilon_i$. Therefore, two conditions
must hold for the coefficient of tail dependence to be non-zero:
\begin{enumerate}
\item the factor must be intrinsically ``wild'' so that its
distribution is regularly varying;
\item the factor must be sufficiently ``wild'' in its
intrinsic wildness, so that its influence
is not dominated by the idiosyncratic component of the asset.
\end{enumerate}
Then, the amplitude of $\lambda_i^\pm$ is determined by the trade-off
between the relative tail behaviors of the factor and the idiosyncratic noise.

As an example, let us consider that the factor and the idiosyncratic
noise follow Student's distribution with $\nu_Y$ and
$\nu_{\epsilon_i}$ degrees of freedom and scale factor $\sigma_Y$ and
$\sigma_{\epsilon_i}$ respectively. Expression (\ref{eq:lambda_add}) leads to
\bea
\lambda_i =& 0 ~&~ \mbox{\rm if} ~~ \nu_Y > \nu_{\epsilon_i}~, \nonumber\\
\label{eq:SFM}
\lambda_i =& \frac{1}{1+ \left( \frac{\sigma_{\epsilon_i}}{\beta_i \cdot
       \sigma_Y} \right)^\nu}~ ~&~ \mbox{\rm if} ~~ \nu_Y =
\nu_{\epsilon_i} = \nu,\\
\lambda_i =& 1 ~&~ \mbox{\rm if} ~~ \nu_Y < \nu_{\epsilon_i}~. \nonumber
\eea
The tail dependence decreases when the
idiosyncratic volatility increases relative to the factor
volatility. Therefore, $\lambda_i$ decreases in period of high
idiosyncratic volatility
and increases in period of high market volatility. From the
viewpoint of the tail dependence, the volatility of an asset is not relevant.
What is governing extreme co-movement is the relative
weights of the different components of the volatility of the asset.

\begin{figure}[h]
\includegraphics[width=15cm]{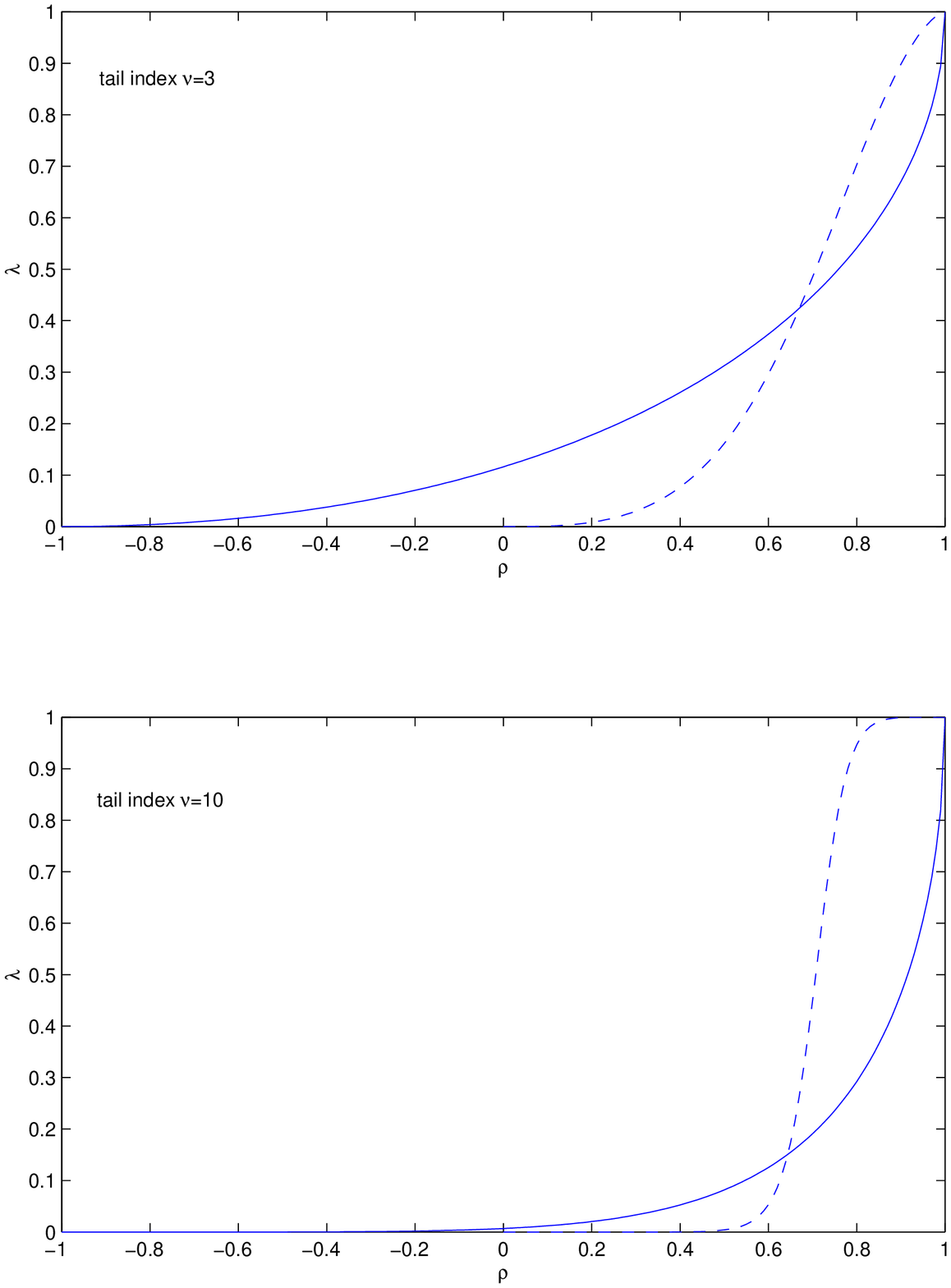}
\caption{\label{fig:1} Evolution as a function of the
correlation coefficient $\rho$ of the coefficient of tail dependence for an
elliptical bivariate Student's distribution (solid line) and for the additive
factor model with Student's factor and noise (dashed line).}
\end{figure}

Figure \ref{fig:1} compares the
coefficient of tail dependence
as a function of the correlation coefficient for
the bivariate Student's distribution (expression (\ref{eq:lambda_mul}))
and for the factor model with the factor and the idiosyncratic noise following
student's distributions (equation (\ref{eq:SFM})).
Contrary to the coefficient of tail
dependence of the Student's factor model, the tail dependence of the elliptical
Student's distribution does not vanish for negative correlation
coefficients. For large values of the correlation coefficient,
the former is always larger than the latter.

Once the coefficients of tail dependence between the assets
and the common factor are known, the
coefficient of tail dependence between any two assets
$X_i$ and $X_j$ with a common factor $Y$ is simply equal
to the weakest tail dependence between the assets and their common factor:
\be
\label{eq:lambda_add2}
\lambda_{ij} = \min \{\lambda_i, \lambda_j \}~.
\ee
This result is very intuitive: since the dependence
between the two assets is due to their common factor, this dependence
cannot be stronger than the weakest dependence between each of the
asset and the
factor.

\section{Practical implementation and consequences }

\begin{figure}[h]
\includegraphics[width=15cm]{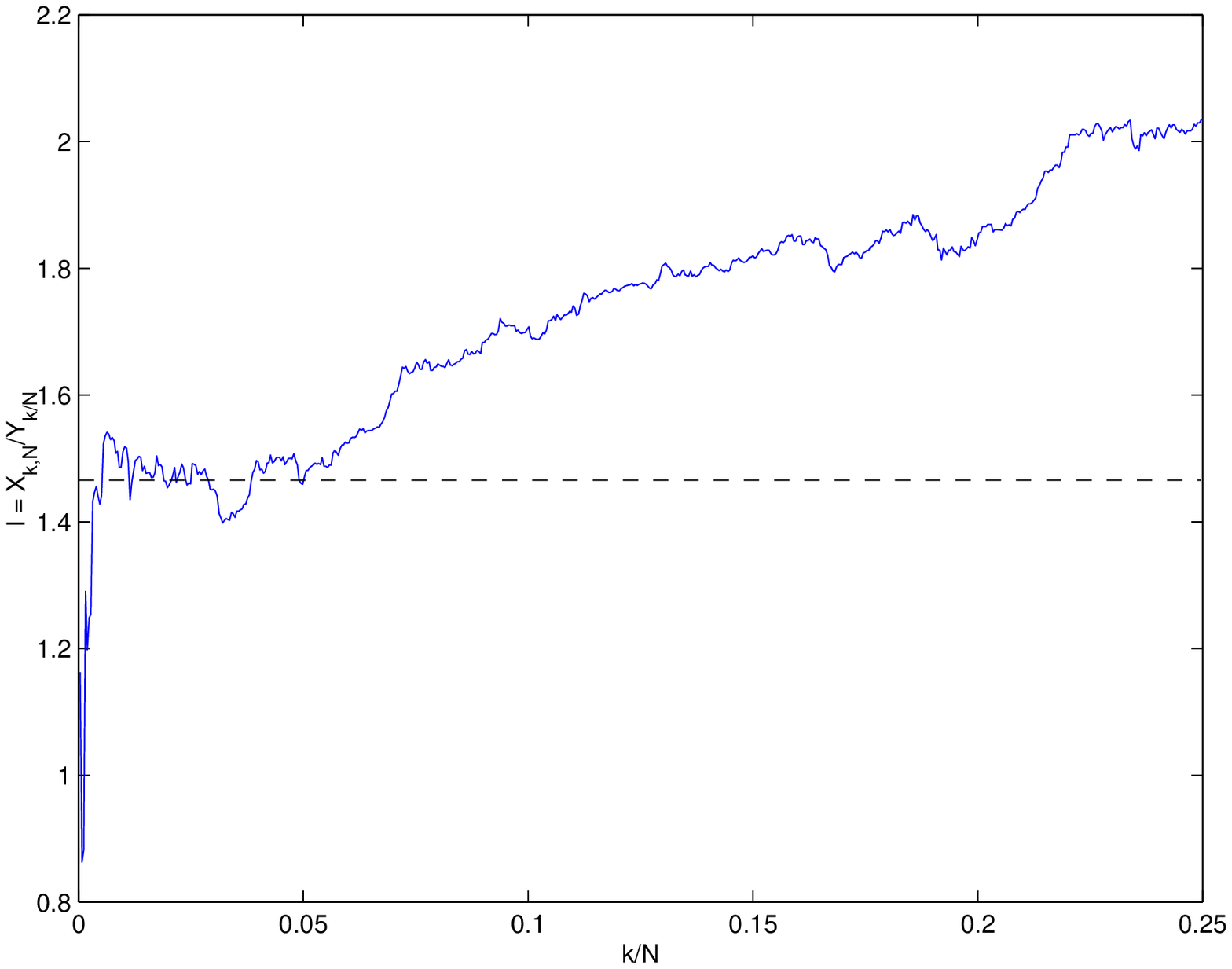}
\caption{\label{fig:2} Empirical estimate $\hat l$ of the
quantile ratio $l$ in (\ref{eq:lambda_mul}) versus the
empirical quantile $\frac{k}{N}$. We observe a very good stabitility
of $\hat l$ for quantiles ranging between 0.005 and  0.05.}
\end{figure}

The two mathematical results (\ref{eq:lambda_mul})
and (\ref{eq:lambda_add}) have a very
important practical impact for the estimation of the
coefficient of tail dependence. As we already pointed out,
its direct estimation is essentially impossible
since the number of
observations goes to zero by definition as the quantile goes to zero.
In constrast, the
formulas (\ref{eq:lambda_mul}) and
(\ref{eq:lambda_add}-\ref{eq:lambda_add2}) tell us that one just has
to estimate a
tail index and a correlation coefficient. These
estimations can be reasonably accurate
because they make use of a significant part of the data beyond the
few extremes targeted by $\lambda$.
Moreover, equation (\ref{eq:lambda_add}) does not explicitly assume a
power law behavior, but only a regularly varying behavior which is far
more general. In such a case, the empirical quantile ratio $l$ in
(\ref{eq:lambda_mul})
turns out to be stable enough for its accurate
non-parametrically estimation, as shown in figure \ref{fig:2}.

\begin{table}[h]
\begin{center}
\begin{tabular}{lcc}
\hline
  & Lower Tail Dependence & Upper Tail Dependence \\
\hline
Bristol-Myers Squibb Co. & 0.16   ~~~( 0.03) & 0.14   ~~~( 0.01) \\
Chevron Corp. & 0.05   ~~~( 0.01) & 0.03   ~~~( 0.01) \\
Hewlett-Packard Co. & 0.13   ~~~( 0.01) & 0.12   ~~~( 0.01) \\
Coca-Cola Co. & 0.12   ~~~( 0.01) & 0.09   ~~~( 0.01) \\
Minnesota Mining \& MFG Co. & 0.07   ~~~( 0.01) & 0.06   ~~~( 0.01) \\
Philip Morris Cos Inc. & 0.04   ~~~( 0.01) & 0.04   ~~~( 0.01) \\
Procter \& Gamble Co. & 0.12   ~~~( 0.02) & 0.09   ~~~( 0.01) \\
Pharmacia Corp. & 0.06   ~~~( 0.01) & 0.04   ~~~( 0.01) \\
Schering-Plough Corp. & 0.12   ~~~( 0.01) & 0.11   ~~~( 0.01) \\
Texaco Inc. & 0.04   ~~~( 0.01) & 0.03   ~~~( 0.01) \\
Texas Instruments Inc. & 0.17   ~~~( 0.02) & 0.12   ~~~( 0.01) \\
Walgreen Co. & 0.11   ~~~( 0.01) & 0.09   ~~~( 0.01) \\
\hline
\end{tabular}
\caption{\label{tab:1} This table presents the coefficients of lower
and of upper
tail dependence with the Standard \& Poor's 500 index
for a set of 12 major stocks traded on de NYSE during the time
interval from January 1991 to December 2000. The numbers within the
parentheses gives the estimated standard deviation of the empirical
coefficients of tail dependence.} \end{center} \end{table}

As an example, the table \ref{tab:1} presents the results obtained
both for the
upper and lower coefficients of tail dependence between several major stocks
and the market factor represented here by the Standard \& Poor's 500 index,
over the last decade. The technical aspects of the method are given in
Appendix \ref{app:1}. The
coefficient of tail dependence between any two assets is easily derived from
(\ref{eq:lambda_add2}).
It is interesting to observe that the coefficients of tail dependence
seem almost identical in the lower and the upper tail. 
  Nonetheless, the coefficient of lower tail dependence is always
  slightly larger than the upper one, showing that large losses are
  more likely to occur together compared with the case of large gains.

\begin{figure}[h]
\includegraphics[width=15cm]{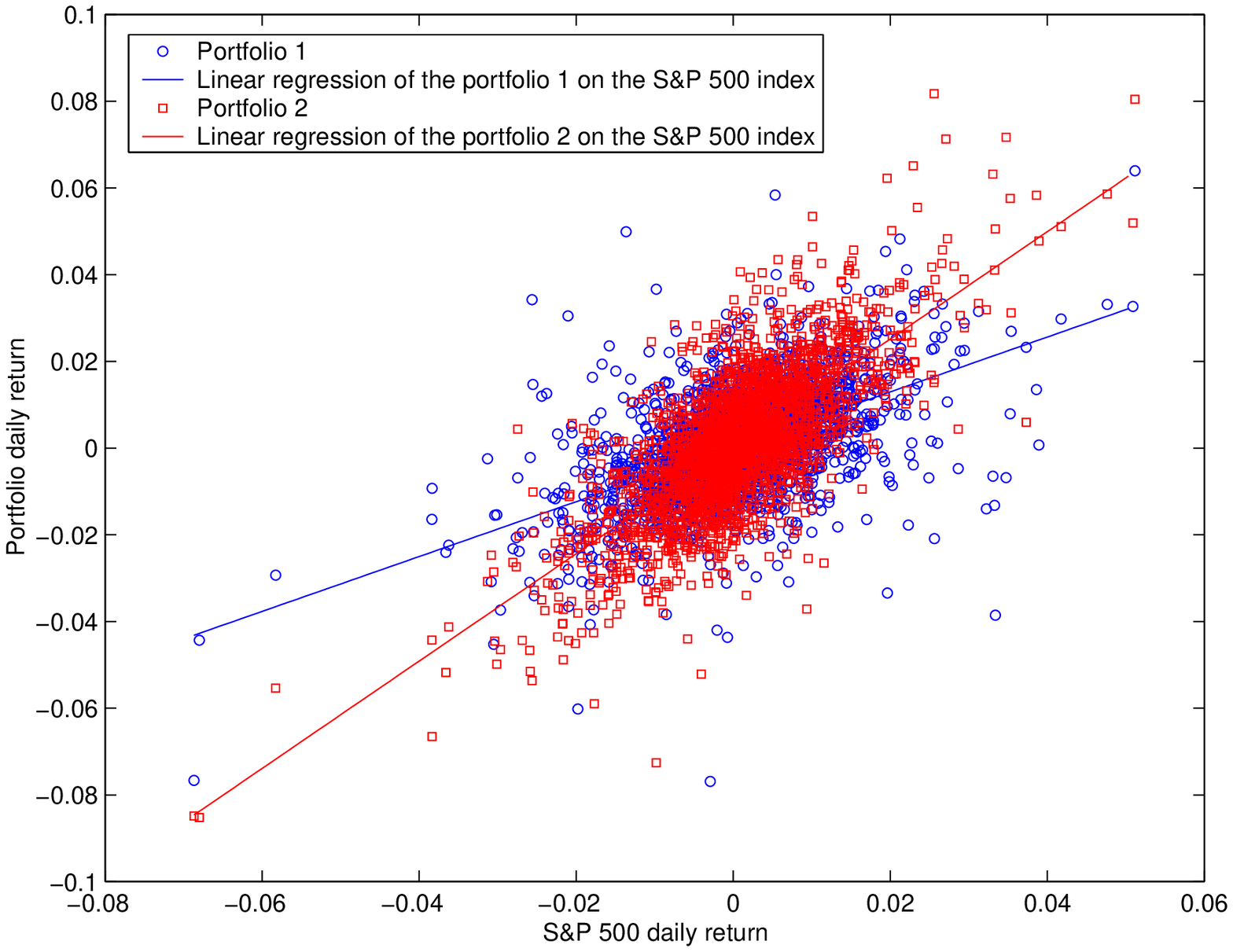}
\caption{\label{benchmark} Daily returns of two equally weighted
portfolios $P_1$ (made of four stocks with small $\lambda \leq 0.06$) and $P_2$
(made of four stocks with large $\lambda \geq 0.12$) as a function
of the daily returns of the Standard \& Poor's 500 over the period
January 1991 to December 2000.
}
\end{figure}

Two clusters of assets clear stand out:
those with a tail dependence of about 10\% (or more) and those with a tail
dependence of about 5\%. These stocks offer the interesting possibility
of devising a prudential portfolio which can be significantly less sensitive
to the large market moves. Figure \ref{benchmark} compares the
daily returns of the Standard \& Poor's 500 index with those of two
portfolios $P_1$ and
$P_2$:  $P_1$ is made of the four stocks (Chevron Corp.,
Philip Morris Cos Inc., Pharmacia Corp. and Texaco Inc.) with the
smallest $\lambda$'s while
$P_2$ is made with the four stocks (Bristol-Meyer Squibb Co.,
  Hewlett-Packard Co., Schering-Plough Corp. and Texas Instruments
  Inc.) with the largest $\lambda$'s. For each set of stocks, we have
constructed
two portfolios, one in which each stock have the same weight $1/4$ and
the other with asset weights
chosen to minimize the variance of the resulting portfolio. We find that
the results are almost the same between the equally-weighted and
minimum-variance
portfolios. This makes sense since extreme tail dependence should not be
controlled by the variance, which accounts only for the price moves
of moderate amplitudes.

Figure \ref{benchmark} presents the results for the equally weighted
portfolios generated from the two groups of assets.
Observe that only one large drop occurs simultaneously for $P_1$ and for
the Standard \& Poor's 500 index in contrast with $P_2$ for which several
large drops are
associated with the largest drops of the index. Quite a few of the
largest drops of $P_2$ occur desynchronized with the index. This
is probably due to the idiosyncratic contributions ${\bf \epsilon}$
in (\ref{idio}) which are in reality not completely independent of the index.
They contain in particular the effect of other factors that have been
left out of this analysis.

Figure \ref{benchmark} shows an almost circular scatter plot for $P_1$ compared
with a rather narrow ellipse for $P_2$: the small tail dependence
between the index and the four stocks in $P_1$ automatically implies that their
mutual tail dependence is also very small, according to (\ref{eq:lambda_add2});
as a consequence, $P_1$ offers a good diversification with respect to large
drops. This effect already quite significant for such small portfolios will be
overwhelming for large ones. The most interesting result stressed in figure
\ref{benchmark} is that optimizing for minimum tail dependence automatically
diversifies away the large risks.

These advantages of portfolio $P_1$ with small tail dependence compared
to portfolio $P_2$ with large tail dependence with the Standard \&
Poor's 500 index  comes at almost no cost in terms of the Sharpe ratio,
equal respectively to 0.058 and 0.061 for the equally weighted and
minimum variance $P_1$ and to 0.069 and 0.071 for the equally weighted and
minimum variance  $P_2$.

The straight lines are bonus: they show that there is
significantly less linear correlations between $P_1$ and the index
(correlation coefficient of $0.52$ for both the equally weigthed and the
minimum variance $P1$) compared with $P_2$ and the index (correlation
coefficient of $0.73$ for the equally weighted $P_2$ and of $0.70$ for the
minimum variance $P_2$). Theoretically, it is possible to
construct two random variables with small correlation coefficient  and
large $\lambda$ and vice-versa. Recall that the correlation
coefficient and the tail dependence
coefficient are two opposite end-members of dependence measures,
the correlation coefficient weighting the dependence between
relatively small moves
while the tail dependence coefficient weighting the dependence
during extreme events. The finding that $P_1$ comes with both
the smallest correlation and the smallest tail dependence
coefficients suggests that there is a continuous of interlaced dependence
structures between assets as a function of the ``depth'' (or quantile) in the
tail of the distribution. This intuition is in fact explained
and encompassed by the factor model since the larger $\beta$ is, the larger is the
correlation coefficient and the larger is the tail dependence. Diversifying away
extreme shocks may provide a useful diversification tool for less extreme
dependences, thus improving the potential usefulness of a strategy of portfolio
management based on tail dependence proposed here.

As a final remark, the almost identical values of the coefficients
of tail dependence for negative and positive tails has the following
consequence: minimizing the
large concomittant losses between the stocks and the market comes with renouncing to
the potential concomittant large gains. This point is well examplified by our
two portfolios (see figure \ref{benchmark}): $P_2$ obviously underwent severe
negative co-movements but it also enjoyed large gains coming together with the large
positive movements of the index. In contrast, $P_1$ is almost completely decoupled
from the
large negative movements of the market but this comes also with its insensitivity
with respect to the large positive movements of the index. Thus, a good dynamical
strategy seems to be the following one: invest in $P_1$ during bearish or
trend-less market phases and prefer $P_2$ in a bullish market.

\appendix

\section{Empirical estimation of the coefficient of tail dependence}
\label{app:1}

This appendix shows how to estimate the coefficient of tail dependence between
an asset $X$ and the market factor $Y$ related by the relation
(\ref{idio}) where $\epsilon$ is an idiosyncratic noise uncorrelated with $X$.

Given a sample of $N$ realizations $\{X_1, X_2, \cdots, X_N \}$ and
$\{Y_1, Y_2, \cdots, Y_N\}$ of $X$ and $Y$, we first estimate the
   coefficient $\beta$ using the ordinary least
square estimator. Let $\hat \beta$ denote its estimate.
Then, using Hill's estimator, we obtain the tail index $\hat \nu$ of
the factor $Y$:
\be
\hat \nu_k = \left[ \frac{1}{k} \sum_{j=1}^k \log Y_{j,N} - \log
   Y_{k,N} \right]^{-1},
\ee
where $Y_{1,N} \ge Y_{2,N} \ge \cdots \ge Y_{N,N}$ is the ordered
statistics of the $N$ realizations of $Y$. The constant $l$ is
non-parametrically estimated with the formula
\be
l = \lim_{u \rightarrow 1} \frac{F_X^{-1}(u)}{F_Y^{-1}(u)} \simeq
\frac{X_{k,N}}{Y_{k,N}},
\ee
for $k =o(N)$, which means that $k$ must remain very small with
respect to $N$ but large enough to ensure an accurate determination of
$l$. The figure \ref{fig:2} presents $\hat l$ as a function of $k/N$.

Finally, using equation (\ref{eq:lambda_add}), the estimated $\hat
\lambda$ is
\be
\hat \lambda^+ = \frac{1}{\max\left\{1, \frac{\hat l}{\hat
       \beta}\right\}^{\hat \nu}}.
\ee


\end{document}